\documentclass[preprintnumbers,prd,twocolumn,showpacs,floatfix,preprintnumbers,superscriptaddress,nofootinbib]{revtex4-2}

\usepackage{graphicx}
\usepackage{epsfig}
\usepackage{bm}
\usepackage{amssymb}
\usepackage{float}
\usepackage{amsmath}
\usepackage{subfigure}
\usepackage{dcolumn}
\allowdisplaybreaks
\usepackage{multirow}
\usepackage[colorlinks]{hyperref}
\usepackage[usenames,dvipsnames]{color}
\hypersetup{
     breaklinks=true,
    pdfstartview={FitH},    
    colorlinks=true,       
    linkcolor=blue,          
    citecolor=red,        
    filecolor=magenta,      
    urlcolor=blue,           
    anchorcolor=green,      
    linktocpage=true
}
\usepackage{orcidlink}

\newcommand{\Mpl}{M_{\textrm{Pl}}}

\newcommand{\nn}{\nonumber}

\def\lam{\lambda}

\def\doi{http://doi.org}

 \def\e{\mathrm{e}}

\allowdisplaybreaks

\begin{document}

\title{Observational constraints on early time non-phantom behaviour of dynamical dark energy}

\author{Sk. Sohail\orcidlink{0009-0003-1839-6816}}
\email{sksohail0402@gmail.com}
\affiliation{Department of Physics, Jamia Millia Islamia, New Delhi, 110025, India}

\author{Sonej Alam\orcidlink{0009-0008-8322-2923}}
\email{sonejalam36@gmail.com}
\affiliation{Department of Physics, Jamia Millia Islamia, New Delhi, 110025, India}

\author{Md. Wali Hossain\orcidlink{0000-0001-6969-8716}}
\email{mhossain@jmi.ac.in}
\affiliation{Department of Physics, Jamia Millia Islamia, New Delhi, 110025, India}


\begin{abstract}
We investigate dynamical dark energy models in which the scalar field component is non-phantom at early times. The models include a single-exponential thawing field, a double-exponential scaling--thawing field, and extensions in which a steep exponential scaling field is combined with a cosmological constant, a constant-$w$ fluid, or a CPL fluid. We confront them with Planck PR4 CMB temperature, polarisation and lensing data, DESI DR2 BAO measurements and the Pantheon$+$ supernova sample. The background parameters are stable across all models, with $H_0 \simeq 67.7$--$68.3\,\mathrm{km\,s^{-1}\,Mpc^{-1}}$, so none of the models relieves the Hubble tension. The CPL parametrization prefers $w_0>-1$ and $w_a<0$ at $\sim\!2.7\sigma$ relative to $\Lambda$CDM, corresponding to phantom behaviour at low redshift. The constant-$w$ and thawing models remain consistent with a cosmological constant. When an early scaling phase is added, the data limit the peak early dark energy fraction to $f_{\mathrm{EDE}}<0.0075$ ($f_{\mathrm{EDE}}<0.015$ for SC$+$CPL). This requires a steep potential, $\lambda_2 \gtrsim 30$, and keeps the scalar-field density below the half-percent level at matter--radiation equality. Early scaling dynamics therefore cannot alter the sound horizon enough to address the Hubble tension. The scaling models fit the data as well as $\Lambda$CDM ($\Delta\chi^2_{\mathrm{min}}\lesssim 1.4$), but they are disfavoured by AIC and BIC because of their additional, weakly constrained parameters. Adding a scaling field to CPL does not improve the fit. Late-time evolving dark energy is mildly preferred by the data, while a simple early scaling phase offers no benefit.
\end{abstract}

\maketitle

\flushbottom

\section{Introduction}
\label{sec:intro}

The discovery of the late time accelerated expansion of the Universe reshaped modern cosmology, pointing to the existence of an unknown component dubbed dark energy~\cite{Copeland:2006wr}. The standard cosmological model, $\Lambda$ Cold Dark Matter ($\Lambda$CDM), attributes this acceleration to the cosmological constant (CC)~\cite{Martin:2012bt,Sahni:1999gb}, which provides an excellent fit to a wide range of observations~\cite{Planck:2018vyg}. However, recent local measurements of the Hubble constant ($H_0$) by the Supernovae and $H_0$ for the Equation of State (EoS) of Dark Energy (SH0ES) collaboration exhibit a $\sim5\sigma$ tension with the value inferred from cosmic microwave background (CMB) observations under the $\Lambda$CDM framework~\cite{Riess:2021jrx}. In addition, the model faces the so called $S_8$ tension associated with the amplitude of matter clustering~\cite{Perivolaropoulos:2021jda,Kilo-DegreeSurvey:2023gfr}.Several other local determinations of $H_0$, however, are consistent with the CMB value~\cite{Boruah:2020fhl,Freedman:2021ahq,Wu:2021jyk,Cao:2022ugh,Chen:2024gnu,Freedman:2024eph}. The $S_8$ tension has also become less significant in recent analyses~\cite{Stolzner:2025htz}.

Furthermore, the recently released Dark Energy Spectroscopic Instrument (DESI) data appear to favour a dynamical dark energy (DDE) component over a pure cosmological constant. The first data release (DR1) of DESI reported a $2.6\sigma$ \cite{DESI:2024mwx} preference for DDE over $\Lambda$CDM in the combined DESI$+$CMB analysis using the Chevallier--Polarski--Linder (CPL) parametrization~\cite{Chevallier:2000qy,Linder:2002et}. This preference increased to $3.1\sigma$ in the second data release (DR2)~\cite{DESI:2025zgx,DESI:2025wyn}. The inclusion of various Type~Ia supernova (SNe~Ia) samples further strengthens this evidence, yielding a $2.8\sigma$--$4.2\sigma$ preference for DDE over the cosmological constant~\cite{DESI:2025zgx,DESI:2025wyn,DESI:2024aqx,DESI:2024kob}. Taken together, these discrepancies indicate possible limitations of $\Lambda$CDM and motivate the exploration of alternative frameworks capable of describing a time-evolving dark energy sector~\cite{Gomez-Valent:2021cbe,Gomez-Valent:2022bku,Giare:2024smz,Berghaus:2024kra,Wolf:2025jlc,Lee:2025pzo,Zhong:2025gyn,Yao:2025wlx,Qu:2024lpx,Wang:2024dka,Giare:2024gpk,Gialamas:2024lyw,Shlivko:2024llw,Ye:2024ywg,Bhattacharya:2024hep,Ramadan:2024kmn,Jiang:2024xnu,Payeur:2024dnq,Malekjani:2024bgi,Wolf:2023uno,Wolf:2024eph,Wolf:2024stt,Park:2024vrw,Dinda:2024kjf,Dinda:2024ktd,Jiang:2024viw,Colgain:2024xqj,Bhattacharya:2024kxp,Akthar:2024tua,Chan-GyungPark:2025cri,Ferrari:2025egk,Wolf:2025jlc,Peng:2025nez,Colgain:2024mtg,Mukherjee:2024ryz,Mukherjee:2025myk,Mukherjee:2025ytj,Berbig:2024aee,Li:2024qso,Li:2024qus,Li:2025owk,Borghetto:2025jrk,Carloni:2024zpl,Liu:2025mub,Wolf:2025jed,Chudaykin:2024gol,Sohail:2024oki,Hossain:2025grx,Cheng:2025lod,Scherer:2025esj,Pedrotti:2025ccw,Escamilla:2023oce,Liu:2025myr,Lu:2025gki,Wolf:2025acj,Cline:2025sbt,Wang:2025dtk,VanRaamsdonk:2025wvj,Peng:2025tqt,Yang:2025oax,Jiang:2025hco,Ishak:2025cay,Shlivko:2025fgv,Silva:2025twg,Gialamas:2025pwv,Cortes:2025joz,Artola:2025zzb,Dinda:2025hiu,Park:2025fbl,Blanco:2025vva,Li:2025muv,Zhang:2025dwu,Alam:2025epg,Hossain:2025gpr,Nojiri:2025low,Nojiri:2025uew,2025IJMPD..3450058P}. Hints of DDE were already present before the DESI measurements, at a comparable level of significance~\cite{Cao:2023eja,Dong:2023jtk,VanRaamsdonk:2023ion,Rubin:2023jdq,deCruzPerez:2024shj,Park:2024vrw,Yang:2024quintom,Yang:2025modified,Halder:2024phantom}.

The simplest one and two parameter extensions of the $\Lambda$CDM model to a DDE scenario are the constant EoS model ($w$CDM) and the CPL parametrization~\cite{Chevallier:2000qy,Linder:2002et}, respectively. While these parametrizations can successfully describe the late time cosmic acceleration, they become unreliable at higher redshifts~\cite{Akthar:2024tua}. To capture the early time behaviour, we therefore introduce a minimally coupled scalar field with a steep exponential potential, which naturally yields a scaling (SC) solution~\cite{Ratra:1987rm,Copeland:1997et} wherein the scalar field energy density tracks the background density while remaining subdominant during the radiation and matter dominated epochs. As the Hubble friction weakens, the field evolves along this scaling trajectory, maintaining an energy density proportional to that of the dominant background component. This intermediate scaling regime prevents premature domination of the scalar field. The scaling behaviour governed by a steep exponential potential is an attractor solution, and thus the field remains in the scaling regime unless a modification in the effective potential occurs at late times~\cite{Akthar:2024tua,Hossain:2014xha,Geng:2015fla}. 

To realise viable late time cosmological dynamics, we combine this early time scaling behaviour with various late time dark energy descriptions to explore the full dynamical evolution of the Universe. Specifically, we consider the following late time scenarios: (i) the $\Lambda$CDM model, (ii) the $w$CDM model, (iii) the CPL parametrization, and (iv) a non-phantom thawing-like behaviour governed by a shallow exponential potential. In addition, we also examine a pure thawing model described by an exponential potential with a smaller slope, operating throughout the entire expansion history.

Parameter estimation for all models is carried out using the Markov Chain Monte Carlo (MCMC) technique, as implemented in \texttt{Cobaya}~\cite{Torrado:2020dgo}, with theoretical predictions computed using a modified version of the Boltzmann code \texttt{CLASS}~\cite{Lesgourgues:2011re,Blas:2011rf}. We employ a combination of the following cosmological data set: the \textit{Planck}~2018 low~$\ell$ TT and EE likelihoods~\cite{Planck:2018vyg,Planck:2019nip}, the \textit{Planck} PR4 high~$\ell$ \texttt{CamSpec} TT, TE, EE likelihood~\cite{Rosenberg:2022sdy} together with the PR4 CMB lensing likelihood~\cite{Carron:2022eyg}, the DESI~DR2 baryon acoustic oscillation (BAO) measurements~\cite{DESI:2025zgx} and the Pantheon$+$ Type~Ia supernova (SNe~Ia) sample~\cite{Scolnic:2021amr,Brout:2022vxf}. To compare the relative performance of different models, we use the Akaike Information Criterion (AIC)~\cite{Akaike74}, the Bayesian Information Criterion (BIC)~\cite{Schwarz:1978tpv} and the Deviance Information Criterion (DIC)~\cite{Spiegelhalter:2002yvw}.

\section{Non-phantom scalar field dynamics}
\label{sec:dyn}

The scalar field dynamics is governed by the Klein--Gordon equation, which in the flat Friedmann-Lema\^{\i}tre-Robertson-Walker (FLRW) metric takes the following form,
\begin{eqnarray}
    \ddot{\phi} + 3 H \dot{\phi} + V'(\phi) = 0 \, ,
    \label{eq:KG}
\end{eqnarray}
where $V(\phi)$ is the scalar field potential and a prime denotes the derivative with respect to the field $\phi$.

As Eq.~\eqref{eq:KG} shows, the scalar field evolution is driven by $V'(\phi)$ rather than the absolute value of $V(\phi)$; it is therefore convenient to introduce dimensionless combinations built from the first and second derivatives of the potential. A standard choice is
\begin{eqnarray}
    \lambda(\phi) &=& -\frac{M_{\rm Pl}\,V'(\phi)}{V(\phi)} \,, \label{eq:lambda_def}\\[4pt]
    \Gamma(\phi) &=& \frac{V(\phi)\,V''(\phi)}{V'(\phi)^2} \,, \label{eq:Gamma_def}
\end{eqnarray}
where $M_{\rm Pl}$ is the reduced Planck mass. Both $\lambda$ and $\Gamma$ are dimensionless: $\lambda$ measures the steepness of the potential in units of the potential itself, while $\Gamma$ quantifies the curvature of the potential relative to the square of its slope.

The evolution of $\lambda$ is governed by $\Gamma(\phi)$ through 
\begin{eqnarray}
    \frac{d\lambda}{dN} 
    = -\sqrt{6}\,x\,\lambda^2(\phi)\,\big[\Gamma(\phi)-1\big]\,, 
    \label{eq:dlambda_dN}
\end{eqnarray}
where $x=\dot{\phi}/(\sqrt{6}\,H\,M_{\rm Pl})$ and $N=\ln a$ with $a$ being the scale factor. If $\Gamma(\phi)=1$, $\lambda$ remains constant, corresponding to the exponential potential $V(\phi)\propto e^{-\lambda\phi/M_{\rm Pl}}$. In this case, the system becomes completely autonomous with constant $\lambda$~\cite{Copeland:1997et}. For general potentials with $\Gamma(\phi)\neq1$, $\lambda$ evolves according to Eq.~\eqref{eq:dlambda_dN}, thereby modifying the dynamics.

Hence, the pair $\{\lambda(\phi),\Gamma(\phi)\}$ fully encapsulates the scalar field dynamics. While the specific potential $V(\phi)$ determines the functional forms of $\lambda(\phi)$ and $\Gamma(\phi)$, it is possible for different potentials to yield similar dependences of these functions on $\phi$. Consequently, such potentials can lead to identical cosmological evolution. This demonstrates that it is the behaviour of $\lambda(\phi)$ and $\Gamma(\phi)$, rather than the explicit form of $V(\phi)$ itself, that fundamentally governs the scalar field dynamics~\cite{Hossain:2023lxs}.

The quantities $\lambda(\phi)$ and $\Gamma(\phi)$ not only characterise the potential but also determine the qualitative behaviour of the scalar field, leading to distinct dynamical classes such as thawing, scaling, and tracker models.

In thawing models~\cite{Scherrer:2007pu}, the scalar field is initially frozen due to strong Hubble friction, i.e. $\dot{\phi}\simeq 0$, and begins to evolve only when $H$ becomes comparable to the effective mass of the field. These models typically correspond to potentials with small $\lambda(\phi)$ at early times, leading to an EoS $w_\phi\simeq -1$ that gradually increases towards higher values as the field rolls down the potential.

In contrast, scaling models~\cite{Copeland:1997et,Hossain:2023lxs} arise when the field energy density evolves proportionally to that of the background fluid, ensuring a constant ratio $\rho_\phi/\rho_{\rm b}$. This behaviour requires $\lambda$ to be large and nearly constant, corresponding to $\Gamma(\phi)\simeq 1$, as realised in exponential potentials $V(\phi)\propto e^{-\lambda\phi/M_{\rm Pl}}$. Such models prevent the scalar field from dominating the energy budget at early times and thus cannot by themselves explain late time cosmic acceleration.

Tracker models, introduced in~\cite{Ratra:1987rm,Peebles:1987ek} and further developed in~\cite{Steinhardt:1999nw,Zlatev:1998tr} (for recent cosmological constraints see~\cite{Hossain:2023lxs,Hossain:2025grx}), represent another important class where the scalar field solution quickly converges to a common evolutionary track regardless of the initial conditions. This behaviour typically arises when $\Gamma(\phi)>1$, which ensures that $\lambda(\phi)$ decreases as the field rolls down the potential. As a result, the field dynamically adjusts to follow the background evolution during the radiation- and matter-dominated eras and naturally transitions to a dark energy–dominated regime at late times.

To capture both early time and late time behaviour of dark energy within a single scalar field model, we consider the double-exponential potential \cite{Barreiro:1999zs}
\begin{equation}
    V(\phi)=V_{1}\e^{-\lambda_{1}\phi/M_{\rm Pl}}+V_{2}\e^{-\lambda_{2}\phi/M_{\rm Pl}}\; ,
    \label{eq:V_double_exp}
\end{equation}
and assume
\begin{equation}
    V_{2}>V_{1}\,,\qquad \lambda_{2}>\lambda_{1}>0,
    \label{eq:hierarchy}
\end{equation}
so that the second term has the larger amplitude and is steeper. With our definitions of $\lambda(\phi)$~\eqref{eq:lambda_def} and
$\Gamma(\phi)$~\eqref{eq:Gamma_def}, a straightforward calculation yields
\begin{equation}
    \Gamma(\phi) = 1-\frac{(\lambda-\lambda_1)(\lambda-\lambda_2)}{\lambda^2}\,.
    \label{eq:Gamma_double}
\end{equation}

\begin{figure}[t]
\centering
\includegraphics[scale=.65]{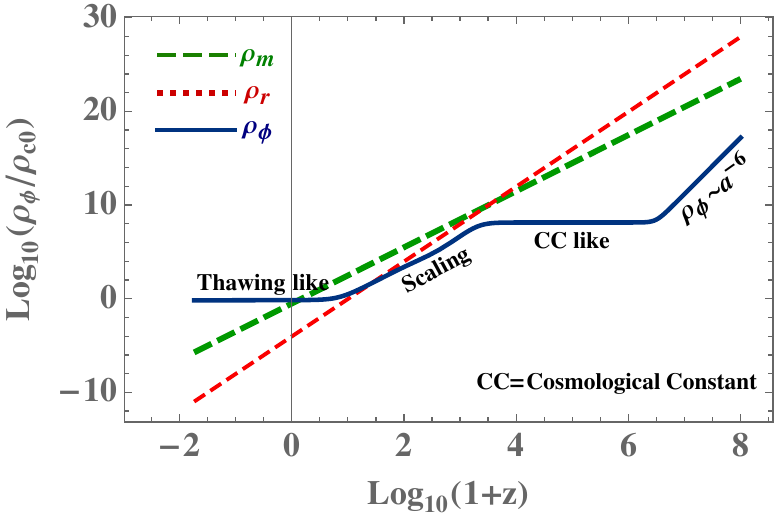} ~~~
\caption{Evolution of the energy densities of matter (long-dashed green), radiation (short-dashed red), and the scalar field (solid blue) normalised by the present critical density $\rho_{\rm c0}$ for the potential~\eqref{eq:V_double_exp} with $\lambda_1=0.1$ and $\lambda_2=20$.}
\label{fig:scaling_dexp}
\end{figure}

Figure~\ref{fig:scaling_dexp} shows the evolution of the scalar field energy density for the potential~\eqref{eq:V_double_exp} along with the matter and radiation energy densities. The different phases of scalar field evolution can be explained using the limits of the function $\Gamma$~\eqref{eq:Gamma_double}.
\noindent
\begin{itemize}
    \item \textbf{Early time behaviour.} As $V_{2}>V_{1}$ and $\lambda_{2}>\lambda_{1}$, for smaller values of $\phi$, if initially
    $V_{2}e^{-\lambda_{2}\phi/M_{\rm Pl}}\gg V_{1}e^{-\lambda_{1}\phi/M_{\rm Pl}}$, then the second term of the potential~\eqref{eq:V_double_exp} dominates. In this limit, $\lambda(\phi)\simeq \lambda_{2}$ and $\Gamma(\phi)\simeq 1$, i.e. the dynamics is approximately that of an exponential potential with slope $\lambda_{2}$. The steepness ($\lambda_2>\sqrt{3}$) prevents early scalar field domination and typically produces scaling-like behaviour~\cite{Ratra:1987rm,Copeland:1997et} (see Fig.~\ref{fig:scaling_dexp}). Before entering the scaling regime, the scalar field remains frozen (CC-like phase in Fig.~\ref{fig:scaling_dexp}) at a higher energy scale ($\sim V_2^{1/4}$) in the past when its effective mass, $m_\phi$, remains smaller than the Hubble scale, $H(t)$, leading to strong Hubble damping. This condition is governed by the second derivative of the scalar field potential, $V(\phi)$, such that  
    \begin{equation}
        m_{\phi}^{2} \equiv V''(\phi) < H^2.
    \end{equation}
    During this frozen phase, the energy density of the scalar field, $\rho_{\phi}$, gradually increases and eventually becomes comparable to the background energy density. When $m_{\phi} \sim H$, the Hubble friction weakens, allowing the scalar field to roll down its potential and enter the scaling regime. Before reaching the CC-like phase, the scalar field energy can also be dominated by the kinetic term due to the large value of $\lambda_2$, and fall as $a^{-6}$, as shown in Fig.~\ref{fig:scaling_dexp}.

    \item \textbf{late time behaviour.} As the field evolves (for example, if $\phi$ increases so that the steeper exponential term becomes subdominant), the term with the smaller slope $\lambda_{1}$ eventually dominates, {\it i.e.},
    $V_{1}e^{-\lambda_{1}\phi/M_{\rm Pl}}\gg V_{2}e^{-\lambda_{2}\phi/M_{\rm Pl}}$. In this regime, $\lambda(\phi)\simeq \lambda_{1}$ and $\Gamma(\phi)\simeq 1$, and the late time evolution is governed by the shallower exponential, suitable for producing slow-roll or near-CC behaviour if $|\lambda_{1}|$ is sufficiently small. This leads to thawing-like behaviour for $\lambda_1<\sqrt{3}$~\cite{Copeland:1997et,Scherrer:2007pu}, as shown in Fig.~\ref{fig:scaling_dexp}.

    \item \textbf{Transition regime.} During the crossover between the two terms, $\lambda(\phi)$ interpolates smoothly between $\lambda_{2}$ and $\lambda_{1}$, while $\Gamma(\phi)$ deviates from unity. During this transition $\lambda_{1}<\lambda(\phi)<\lambda_{2}$, and from Eq.~\eqref{eq:Gamma_double} we see that $\Gamma(\phi)>1$. This implies that the field may briefly exhibit tracker-like behaviour during the transition, eventually evolving rapidly from a scaling regime to a potential-dominated thawing-like phase. The duration and nature of this transition depend sensitively on the relative magnitudes of $\lambda_{1}$, $\lambda_{2}$, and the ratio $V_{2}/V_{1}$, which together determine whether the evolution is smooth or features a pronounced intermediate scaling epoch.
\end{itemize}

Because we employ a canonical kinetic term, the scalar field remains non-phantom throughout the evolution, i.e. for a canonical field one always has $w_{\phi} \geq -1$. However, the model can be extended to accommodate late time phantom behaviour by introducing an appropriate parametrization for the late time evolution of dark energy, which we discuss in the next section. In this work, we focus on scenarios in which the dark energy dynamics is non-phantom during the early Universe and may or may not exhibit phantom behaviour at late times.

\section{Models}
\label{sec:models}

We focus on different classes of dark energy (DE) models. As discussed in Section~\ref{sec:dyn}, a double-exponential potential~\eqref{eq:V_double_exp} can give rise to an early time scaling solution followed by a non-phantom thawing dynamics. To explore such behaviour, we consider an exponential potential to study the thawing model and a double-exponential potential to capture the early dynamics of a non-phantom scalar field. Furthermore, we modify the late time behaviour by introducing a cosmological constant (CC) term that drives the present accelerated expansion, replacing the exponential term responsible for the late time thawing behaviour. In addition, we also consider the $w$CDM and CPL parametrizations as alternatives to the CC for describing the late time cosmology. While both the $w$CDM and CPL forms can accommodate phantom behaviour, the CPL parametrization offers greater flexibility as it can describe transient phantom dynamics, wherein the EoS temporarily crosses the phantom divide ($w=-1$) before returning to non-phantom values. 

For all models discussed below, we impose the flatness condition
\begin{equation}
\Omega_{\rm m0} + \Omega_{\rm r0} + \Omega_{\rm DE0} = 1,
\label{eq:flatness_condition}
\end{equation}
where $\Omega_{\rm m0}$, $\Omega_{\rm r0}$, and $\Omega_{\rm DE0}$ denote the present-day fractional energy densities of matter, radiation, and dark energy, respectively.

\subsection{Thawing (TH)}
\label{sec:thaw}

We begin by considering the thawing class of scalar field models characterised by an exponential potential of the form
\begin{equation}
V(\phi) = V_{1} e^{-\lambda_{1} \phi / M_{\rm Pl}},
\label{eq:V_exp}
\end{equation}
where $V_{1}$ and $\lambda_{1}$ are positive constants. At early times, when the Hubble damping term dominates over the slope of the potential (Eq.~\eqref{eq:KG}), the scalar field remains nearly frozen due to the large Hubble friction. Consequently, its EoS parameter $w_{\phi}$ stays close to $-1$, mimicking a cosmological constant. As the Universe expands and the Hubble parameter decreases, the friction term weakens and the field begins to ``thaw'' from its frozen state, gradually rolling down the potential.

The subsequent evolution of the field is governed by the steepness parameter $\lambda_{1}$. For $\lambda_{1}<\sqrt{3}$~\cite{Copeland:1997et}, the potential is sufficiently flat, and the field evolves slowly, maintaining $w_{\phi}\simeq-1$ for an extended period, thereby providing a viable explanation for the observed late time acceleration. Larger values of $\lambda_{1}$ correspond to steeper potentials, leading to faster evolution of $\phi$ and larger deviations of $w_{\phi}$ from $-1$. However, since the kinetic term remains canonical, the scalar field always satisfies the non-phantom condition $w_{\phi} \geq -1$.

\subsection{Scaling$+$Thawing (SC$+$TH)}
\label{sec:sc+thaw}

The double-exponential potential~\eqref{eq:V_double_exp} provides a unified framework \cite{Barreiro:1999zs,Ali:2012cv,Hossain:2017ica,Ramadan:2023ivw} in which the scalar field exhibits an early time scaling behaviour followed by a late time thawing phase. The steep exponential term (with slope $\lambda_{2}$) ensures that the field energy density tracks the background at early epochs, avoiding premature domination, while the shallow exponential term (with slope $\lambda_{1}$) becomes relevant at late times, allowing the field to slowly roll and drive cosmic acceleration. This transition from scaling to thawing—termed ``Scaling$+$Thawing''—thus naturally connects early non-dominant dynamics with late time accelerated expansion within a single non-phantom scalar field model. The parameters $\{V_{1},V_{2},\lambda_{1},\lambda_{2}\}$ are constrained such that the total energy density today satisfies the flatness condition~\eqref{eq:flatness_condition}.

\subsection{Scaling$+\Lambda$ (SC$+\Lambda$)}
\label{sec:sc+lambda}

In this variant, we retain the steep exponential term responsible for the early time scaling behaviour but replace the shallow exponential component with a cosmological constant ($\Lambda$) term. The potential takes the form
\begin{equation}
V(\phi)=V_{2}e^{-\lambda_{2}\phi/M_{\rm Pl}} \, ,
\label{eq:V_scaling}
\end{equation}
and the dark energy density is given by
\begin{equation}
\rho_{\rm DE}=\frac{1}{2}\dot \phi^2 + V_{2}e^{-\lambda_{2}\phi/M_{\rm Pl}}+\Lambda.
\label{eq:V_scaling_lambda}
\end{equation}
During the early Universe, the steep exponential ensures that the scalar field energy density scales with the background (radiation or matter), preventing early domination and maintaining consistency with nucleosynthesis and structure formation constraints. As the Universe expands and $H$ decreases, the contribution of the exponential term becomes subdominant compared to the constant term $\Lambda$, which eventually drives the late time acceleration. 

This construction therefore provides a simple non-phantom extension of the standard $\Lambda$CDM scenario that still retains the desirable early time scaling dynamics of quintessence models. The transition from the scaling regime to the $\Lambda$-dominated phase is governed by the relative magnitudes of $V_{2}$, $\Lambda$, and the steepness parameter $\lambda_{2}$. The present-day densities again satisfy the flatness condition~\eqref{eq:flatness_condition}, with $\Omega_{\rm DE0}=\Omega_{\Lambda0}$, the present fractional energy density of the cosmological constant.

\subsection{Scaling$+w$CDM (SC$+w$)}
\label{sec:sc+wcdm}

In this model, the early time dynamics is governed by a canonical scalar field with a steep exponential potential~\eqref{eq:V_scaling},
which gives rise to the desired scaling behaviour during the
radiation and matter dominated eras. The late time dark energy sector is introduced as an independent effective fluid with a constant equation of state parameter $w$. Thus, in SC+$w$ the energy density of dark energy is given by
\begin{equation}
\rho_{\rm DE} = \frac{1}{2}\dot{\phi}^{\,2} + V(\phi) + \rho_{w0}\,a^{-3(1+w)},
\end{equation}

\begin{table*}[t]
\centering
\caption{Constraints on the parameters for the $\Lambda$CDM, Thawing, SC+TH (Double Exponential), SC+$\Lambda$CDM, SC+$w$CDM, SC+CPL and CPL models. Central values are the best-fit values and the error bars are the marginalized $68\%$ ($1\sigma$) intervals. The upper block lists $\Omega_{\rm m0}$, $H_0$, $\omega_{\rm b}$, $r_{\rm d}h$, $f_{\rm EDE}$ and $\log_{10}(1+z_c)$; the lower block lists the remaining parameters for the same models.}
\label{tab1}
\footnotesize
\renewcommand{\arraystretch}{1.3}
\setlength{\tabcolsep}{3pt}
\begin{tabular*}{\textwidth}{@{\extracolsep{\fill}} l c c c c c c}
\hline\hline
Model & $\Omega_{\rm m0}$ & $H_0$ & $\omega_{\rm b}$ & $r_{\rm d}h$ & $f_{\rm EDE}$ & $\log_{10}(1+z_c)$ \\
\hline
$\Lambda$CDM
& $0.3023^{+0.0036}_{-0.0036}$ & $68.26^{+0.2819}_{-0.2785}$ & $0.02242\pm 0.000121$ & $100.8^{+0.4750}_{-0.4691}$ & $\cdots$ & $\cdots$ \\
Thawing
& $0.3053^{+0.0043}_{-0.0055}$ & $67.77^{+0.5954}_{-0.4072}$ & $0.02234 \pm 0.000123$ & $100.2^{+0.8706}_{-0.6233}$ & $\cdots$ & $\cdots$ \\
SC$+$TH
& $0.3027^{+0.0036}_{-0.0035}$ & $68.27^{+0.2844}_{-0.2785}$ & $0.02241^{+0.000121}_{-0.000122}$ & $100.7^{+0.4675}_{-0.4730}$ & $< 0.0075$ & $3.313\pm 0.2564$ \\
SC$+\Lambda$
& $0.3053^{+0.0035}_{-0.0035}$ & $67.95^{+0.2735}_{-0.2701}$ & $0.02222^{+0.000124}_{-0.000125}$ & $100.4^{+0.4580}_{-0.4523}$ & $< 0.0075$ & $3.323 \pm 0.2592$ \\
SC$+w$
& $0.3025^{+0.0050}_{-0.0051}$ & $68.22^{+0.5657}_{-0.5533}$ & $0.02233\pm 0.000126$ & $100.7^{+0.8230}_{-0.7978}$ & $< 0.0075$ & $3.647\pm 0.2310$ \\
SC$+$CPL
& $0.3075^{+0.0058}_{-0.0057}$ & $67.88^{+0.5936}_{-0.5960}$ & $0.02223\pm 0.000126$ & $100.1^{+0.8862}_{-0.8958}$ & $< 0.015$ & $3.317\pm 0.2615$ \\
CPL
& $0.3102^{+0.0057}_{-0.0058}$ & $67.68^{+0.5941}_{-0.6134}$ & $0.02227\pm 0.000126$ & $99.78^{+0.8903}_{-0.9030}$ & $\cdots$ & $\cdots$ \\
\hline\hline
Model & $\lambda_1$ & $\lambda_2$ & $V_1$ & $\log_{10}V_2$ & $w_0$ & $w_a$ \\
\hline
$\Lambda$CDM
& $\cdots$ & $\cdots$ & $\cdots$ & $\cdots$ & $\cdots$ & $\cdots$ \\
Thawing
& $< 0.7$ & $\cdots$ & $0.7260^{+0.0134}_{-0.0369}$ & $\cdots$ & $\cdots$ & $\cdots$ \\
SC$+$TH
& Unconstrained & $60.84^{+11.98}_{-17.22}$ & Unconstrained & $14.068^{+1.044}_{-1.758}$ & $\cdots$ & $\cdots$ \\
SC$+\Lambda$
& $\cdots$ & $67.27^{+12.81}_{-15.16}$ & $\cdots$ & $36.78^{+5.590}_{-6.754}$ & $\cdots$ & $\cdots$ \\
SC$+w$
& $\cdots$ & $50.02^{+13.58}_{-16.69}$ & $\cdots$ & $29.62^{+6.280}_{-7.331}$ & $\cdots$ & $\cdots$ \\
SC$+$CPL
& $\cdots$ & $82.53^{+10.37}_{-18.89}$ & $\cdots$ & $46.57^{+4.757}_{-8.191}$ & $-0.8767^{+0.0549}_{-0.0564}$ & $-0.4943^{+0.2136}_{-0.2108}$ \\
CPL
& $\cdots$ & $\cdots$ & $\cdots$ & $\cdots$ & $-0.8422^{+0.0554}_{-0.0500}$ & $-0.6250^{+0.1956}_{-0.1913}$ \\
\hline\hline
Model & $\omega_{\rm cdm}$ & $100\,\theta_{\rm s}$ & $n_{\rm s}$ & $\ln(10^{10}A_{\rm s})$ & $\tau_{\rm reio}$ \\
\hline
$\Lambda$CDM
& $0.1178^{+0.0006}_{-0.0006}$ & $1.042^{+0.0002}_{-0.0002}$ & $0.9665^{+0.0034}_{-0.0033}$ & $3.048^{+0.0139}_{-0.0138}$ & $0.0537^{+0.0066}_{-0.0072}$ \\
Thawing
& $0.1172^{+0.0006}_{-0.0006}$ & $1.042^{+0.0002}_{-0.0002}$ & $0.9704^{+0.0033}_{-0.0034}$ & $3.064^{+0.0140}_{-0.0141}$ & $0.0643^{+0.0067}_{-0.0075}$ \\
SC$+$TH
& $0.1180^{+0.0006}_{-0.0006}$ & $1.042^{+0.0002}_{-0.0002}$ & $0.9700^{+0.0034}_{-0.0034}$ & $3.053^{+0.0138}_{-0.0142}$ & $0.0563^{+0.0066}_{-0.0074}$ \\
SC$+\Lambda$
& $0.1181^{+0.0006}_{-0.0006}$ & $1.042^{+0.0002}_{-0.0002}$ & $0.9672^{+0.0033}_{-0.0033}$ & $3.056^{+0.0130}_{-0.0151}$ & $0.0578^{+0.0063}_{-0.0077}$ \\
SC$+w$
& $0.1178^{+0.0007}_{-0.0008}$ & $1.042^{+0.0002}_{-0.0002}$ & $0.9675^{+0.0037}_{-0.0033}$ & $3.055^{+0.0141}_{-0.0141}$ & $0.0587^{+0.0068}_{-0.0075}$ \\
SC$+$CPL
& $0.1188^{+0.0008}_{-0.0008}$ & $1.042^{+0.0002}_{-0.0002}$ & $0.9654^{+0.0036}_{-0.0037}$ & $3.029^{+0.0136}_{-0.0136}$ & $0.0465^{+0.0066}_{-0.0072}$ \\
CPL
& $0.1191^{+0.0008}_{-0.0007}$ & $1.042^{+0.0002}_{-0.0002}$ & $0.9640^{+0.0034}_{-0.0035}$ & $3.037^{+0.0142}_{-0.0142}$ & $0.0519^{+0.0068}_{-0.0074}$ \\
\hline\hline
\end{tabular*}
\end{table*}

\begin{table*}[t]
\centering
\caption{Model comparison statistics for the $\Lambda$CDM, Thawing, SC+TH (Double Exponential), SC+$\Lambda$CDM, SC+$w$CDM, SC+CPL and CPL models (Planck PR4 + PR4 lensing + DESI DR2 + Pantheon+). All differences are with respect to $\Lambda$CDM.}
\label{tab:model_comparison}
\footnotesize
\renewcommand{\arraystretch}{1.3}
\setlength{\tabcolsep}{1.5pt}
\begin{tabular*}{\textwidth}{@{\extracolsep{\fill}} c c c c c c c c}
\hline\hline
Stat. & $\Lambda$CDM & Thawing & SC$+$TH & SC$+\Lambda$ & SC$+w$ & SC$+$CPL & CPL \\
\hline\hline
$\chi^2_{\rm min}$ & $12394.34$ & $12392.70$ & $12394.46$ & $12395.51$ & $12395.73$ & $12386.72$ & $12384.41$ \\
$k$                & $15$       & $16$       & $18$       & $17$       & $18$       & $19$       & $17$       \\
$N$                & $11779$    & $11779$    & $11779$    & $11779$    & $11779$    & $11779$    & $11779$    \\
AIC                & $12424.34$ & $12424.70$ & $12430.46$ & $12429.51$ & $12431.73$ & $12424.72$ & $12418.41$ \\
BIC                & $12534.96$ & $12542.69$ & $12563.19$ & $12554.87$ & $12564.47$ & $12564.83$ & $12543.77$ \\
DIC                & $12419.49$ & $12419.21$ & $12424.79$ & $12424.12$ & $12425.60$ & $12416.06$ & $12413.52$ \\
\hline
$\Delta\chi^2_{\rm min}$ & $0.00$ & $-1.64$ & $0.11$  & $1.17$  & $1.39$  & $-7.62$ & $-9.93$ \\
$\Delta$AIC              & $0.00$ & $0.36$  & $6.11$  & $5.17$  & $7.39$  & $0.38$  & $-5.93$ \\
$\Delta$BIC              & $0.00$ & $7.73$  & $28.23$ & $19.92$ & $29.51$ & $29.87$ & $8.82$  \\
$\Delta$DIC              & $0.00$ & $-0.29$ & $5.30$  & $4.62$  & $6.11$  & $-3.43$ & $-5.98$ \\
\hline\hline
\end{tabular*}
\end{table*}

\begin{figure*}[ht]
\centering
\includegraphics[width=0.45\textwidth]{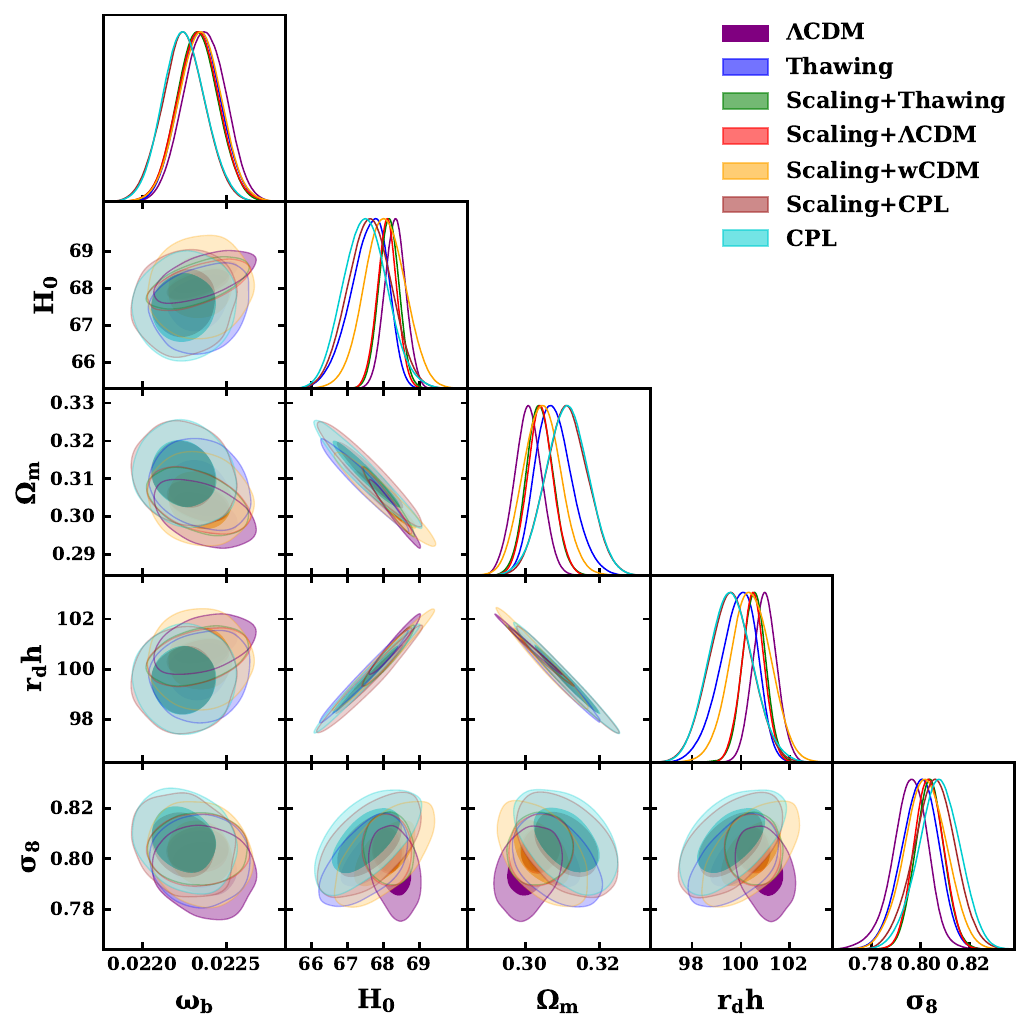}
\hfill
\includegraphics[width=0.45\textwidth]{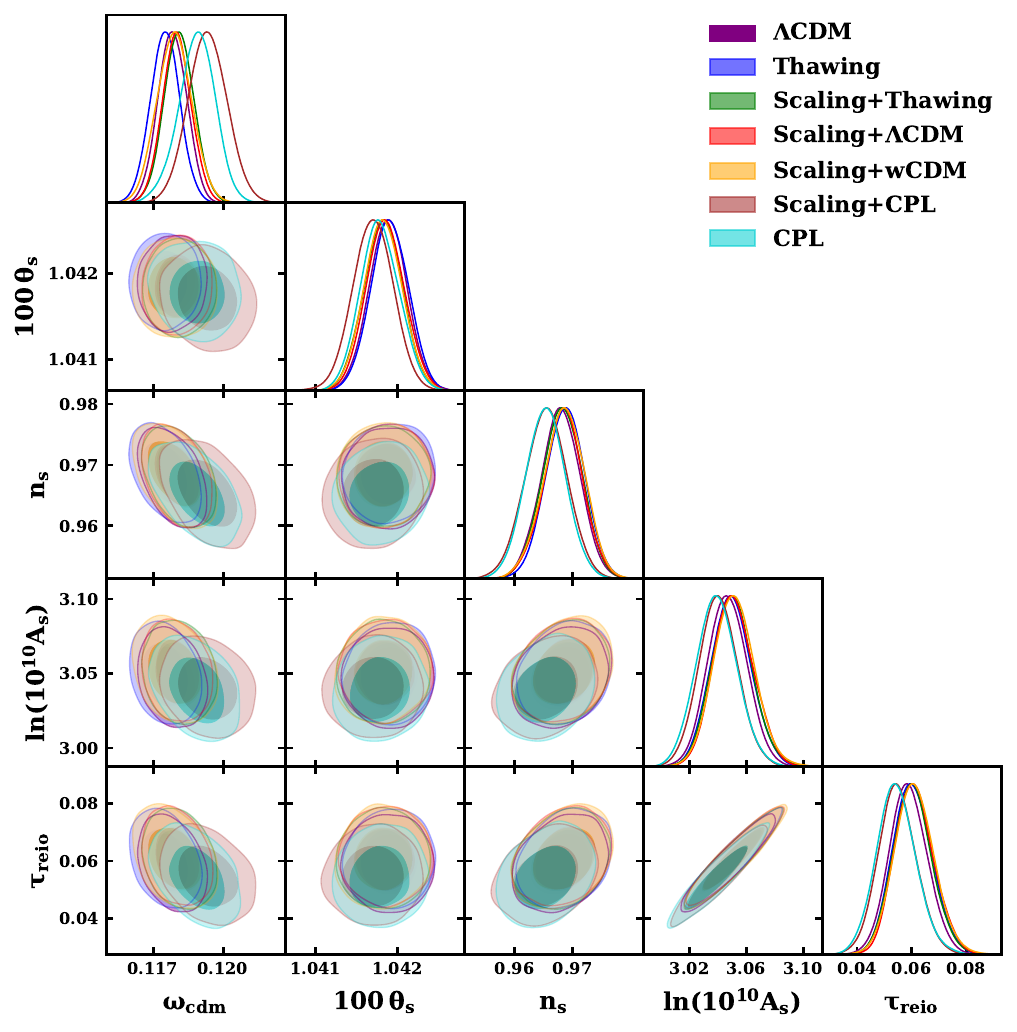}
\caption{The $1\sigma$ and $2\sigma$ confidence regions for the different cosmological parameters in our analysis.}
\label{fig:constraints}
\end{figure*}

\begin{figure*}[ht]
\centering
\includegraphics[width=0.45\textwidth]{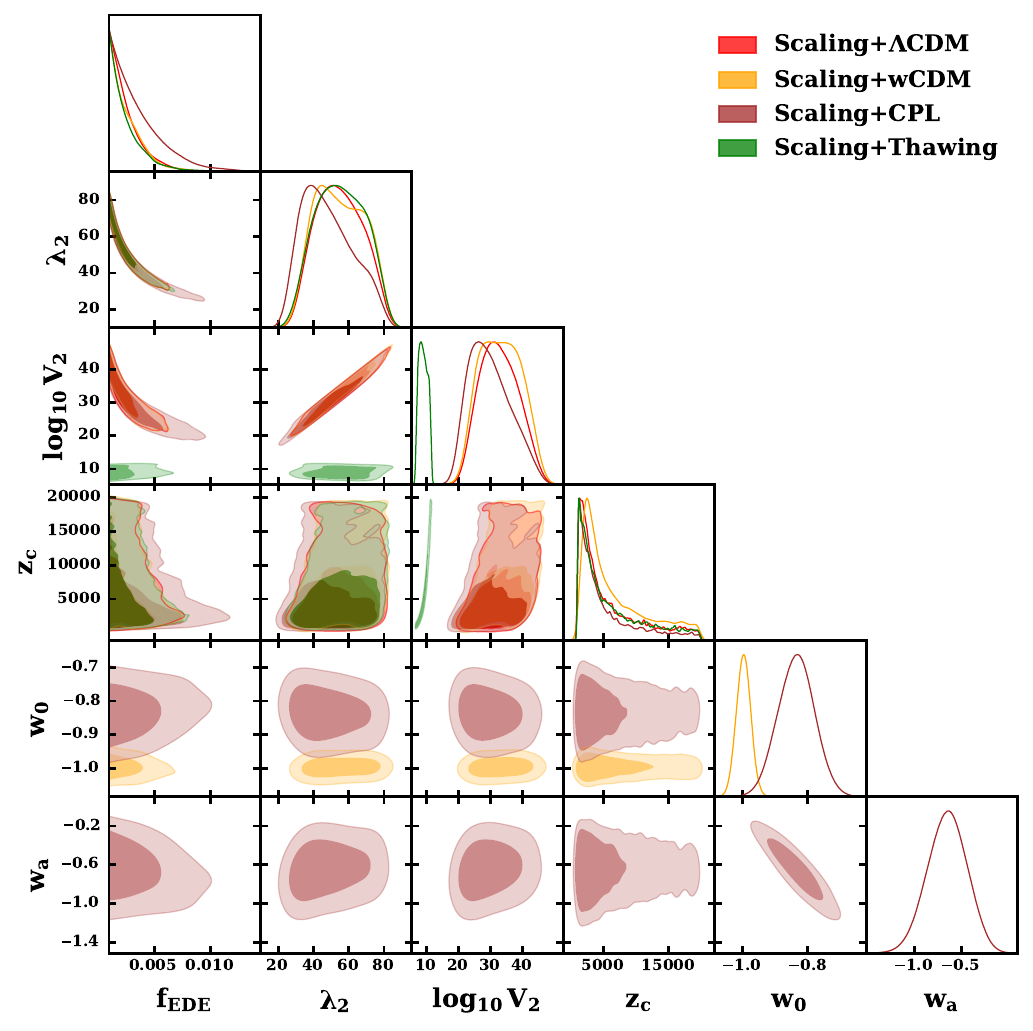}
\caption{The $1\sigma$ and $2\sigma$ confidence regions for the dark energy model parameters.}
\label{fig:dark_scf}
\end{figure*}

\begin{figure}[t]
\centering
\includegraphics[width=0.4\textwidth]{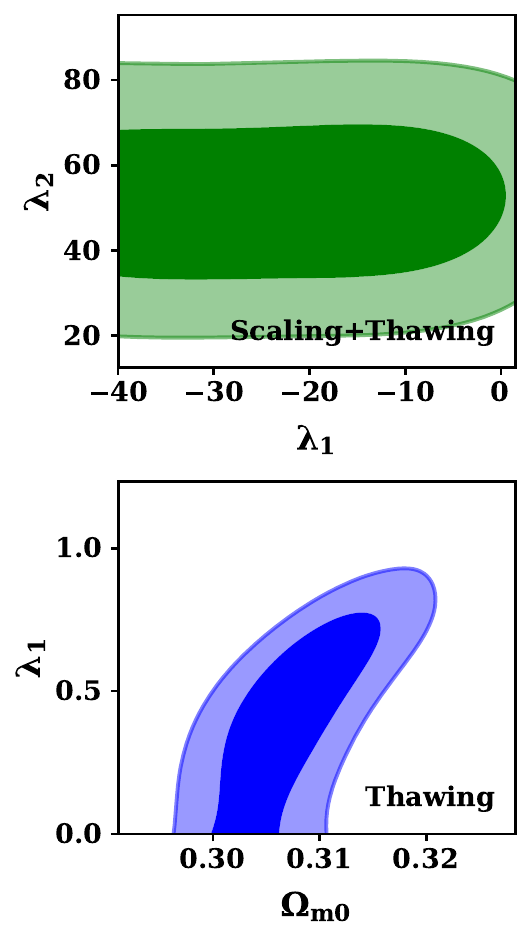}
\caption{Two-dimensional contours in the $(\lambda_1,\lambda_2)$ plane for the scaling--thawing model (upper) and $(\Omega_{\rm m0},\lambda_1)$ plane for the thawing model (bottom).}
\label{fig:constraints_lambda}
\end{figure}

\begin{figure}[t]
\centering
\includegraphics[width=0.42\textwidth]{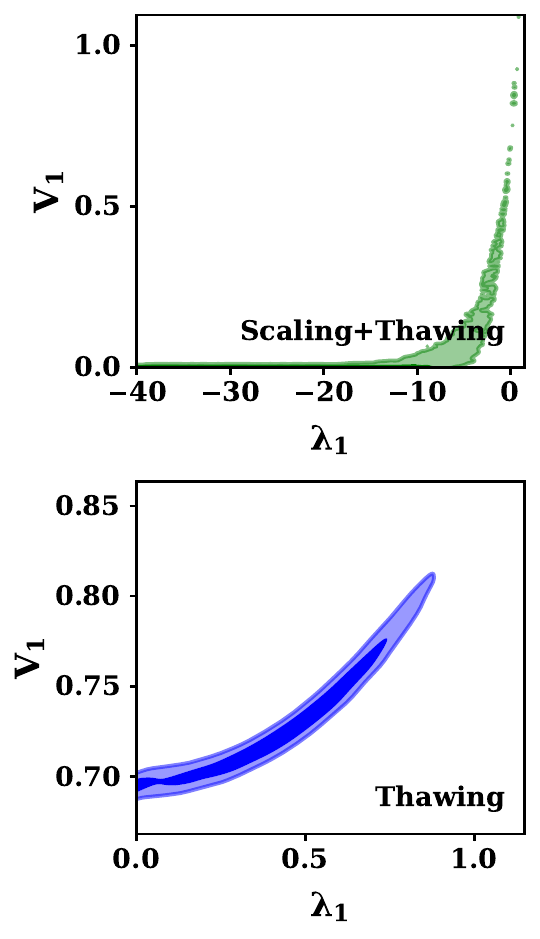}
\caption{The contour in the $\lambda_1$-$V_1$ plane for the Scaling+Thawing (upper) and Thawing (lower) models.}
\label{fig:lambda1_V1}
\end{figure}

\subsection{Scaling$+$CPL (SC+CPL)}
\label{sec:sc+cpl}
Finally, in the ``Scaling$+$CPL'' construction, we generalise the late time dark energy component beyond a constant EoS by allowing it to evolve dynamically according to the Chevallier–Polarski–Linder (CPL) parametrization~\cite{Chevallier:2000qy,Linder:2002et},
\begin{equation}
w(a) = w_{0} + w_{a}(1-a),
\label{eq:w_cpl}
\end{equation}
where $w_{0}$ is the present day equation of state and $w_{a}$ characterises its evolution. The energy density corresponding the CPL parametrization is 
\begin{equation}
\rho_{\rm CPL}(a)
=
\rho_{\rm CPL,0}\,
a^{-3(1+w_{0}+w_{a})}
\exp\left[3w_{a}(a-1)\right].
\label{eq:rho_cpl}
\end{equation}
As in the SC+$w$ case, the scalar field itself has only the steep exponential potential~\eqref{eq:V_scaling} the dark energy density yields
\begin{equation}
\rho_{\rm DE} = \frac{1}{2}\dot{\phi}^{\,2} + V(\phi) + \rho_{\rm CPL},
\end{equation}

\section{Observational data}
\label{sec:data}
We constrain the models with a Markov chain Monte Carlo (MCMC) analysis using the Bayesian sampling framework \texttt{Cobaya}~\cite{Torrado:2020dgo}, with the Metropolis-Hastings sampler and fast-slow parameter decomposition~\cite{Lewis:2013hha}. For each model we have used a modified version of the Boltzmann code \texttt{CLASS} v3.4.0~\cite{Lesgourgues:2011re,Blas:2011rf}, in which we implemented the scalar field potentials and the shooting procedures described below. The non linear matter power spectrum is computed with \texttt{Halofit}~\cite{Takahashi:2012em}, recombination with \texttt{HyRec}~\cite{Ali-Haimoud:2010hou,Lee:2020obi}, and we assume $N_{\rm eff}=3.044$ with one massive neutrino of mass $0.06\,{\rm eV}$. The data combination consists of the Planck 2018 low-$\ell$ TT and EE likelihoods~\cite{Planck:2019nip}, the Planck PR4 (\texttt{NPIPE}) high-$\ell$ \texttt{CamSpec} TT, TE, EE likelihood~\cite{Rosenberg:2022sdy}, the Planck PR4 CMB lensing likelihood~\cite{Carron:2022eyg}, the BAO measurements from the second data release (DR2) of DESI, based on over 14 million extragalactic objects including emission line galaxies, luminous red galaxies, quasars~\cite{DESI:2025qqy} and Lyman-$\alpha$ forest tracers~\cite{DESI:2025zpo}, which provide measurements of $D_M/r_{\rm d}$ and $D_H/r_{\rm d}$ over $0.4<z<4.2$ and of $D_V/r_{\rm d}$ at $0.1<z<0.4$~\cite{DESI:2025zgx} and the Pantheon+ Type Ia supernova sample~\cite{Scolnic:2021amr,Brout:2022vxf} containing 1590 SNe Ia luminosity distance measurements. The chains are considered converged when the Gelman-Rubin statistic satisfies $R-1<0.03$~\cite{Gelman:1992zz}and the posteriors are analysed with \texttt{GetDist}~\cite{Lewis:2019xzd}, discarding the first $30\%$ of each chain as burn in.

\section{ Parameter estimation}
\label{sec:cons}
To evaluate the relative performance of the dark energy models considered in this work, we employ standard model selection criteria in addition to the minimum chi-squared value, $\chi^2_{\rm min}$. In particular, we compute the Akaike Information Criterion (AIC), the Bayesian Information Criterion (BIC) and the Deviance Information Criterion (DIC), defined as
\begin{eqnarray}
{\rm AIC} &=& \chi^2_{\rm min} + 2k, \\
{\rm BIC} &=& \chi^2_{\rm min} + k \ln N, \\
{\rm DIC} &=& 2\langle\chi^2\rangle - \chi^2_{\rm min},
\end{eqnarray}
where $k$ is the number of free parameters, $N$ the total number of data points, and $\langle\chi^2\rangle$ the mean of $\chi^2$ over the posterior samples to assess the differences between the values of each model and those of $\Lambda$CDM. Negative values indicate support for the model over $\Lambda$CDM.

\subsection{Priors}

In all models we sample the six standard cosmological parameters with uniform priors,
\begin{align}
\ln(10^{10}A_{\rm s}) &\in [1.61,\, 3.91], & \quad n_{\rm s} &\in [0.8,\, 1.2], \nn \\
100\,\theta_{\rm s} &\in [0.5,\, 10], & \quad \omega_{\rm b} &\in [0.005,\, 0.1], \nn \\
\omega_{\rm cdm} &\in [0.001,\, 0.99], & \quad \tau_{\rm reio} &\in [0.01,\, 0.8], \nn
\end{align}
where $A_{\rm s}$ and $n_{\rm s}$ are the amplitude and spectral index of the primordial power spectrum respectively,$\,\theta_{\rm s}$ is the angular size of the sound horizon at recombination, $\omega_{\rm b}=\Omega_{\rm b}h^2$ and $\omega_{\rm cdm}=\Omega_{\rm cdm}h^2$ is the physical baryon and cold dark matter densities and $\tau_{\rm reio}$ is the optical depth to reionization. These are sampled together with the nuisance parameters of the Planck likelihoods, with their default priors.The Hubble constant is constrained by shooting on $100\,\theta_{\rm s}$ in \texttt{CLASS}, while $\Omega_{\rm m0}$, $r_{\rm d}h$ and $\sigma_8$ are derived parameters. These priors are common to $\Lambda$CDM and to all extended models.
For the early dark energy models we additionally sample the maximum fractional contribution of early dark energy to the total energy density, $f_{\rm EDE}$ and the redshift $z_c$ at
which this maximum is reached with uniform priors $f_{\rm EDE} \in [0.001,\, 0.3]$ and $\log_{10}(1+z_c) \in [3.0,\, 4.3]$. For the exponential potential, a change in the initial
field value $\phi_{\rm ini}$ is exactly absorbed into $V_2$ by the shooting procedure. Hence, the choice of $\phi_{\rm ini}$ does not affect the constraints on $f_{\rm EDE}$ and $z_c$.The model specific parameters with uniform priors are:
\begin{itemize}
    \item \textbf{Thawing model} (Subsec.~\ref{sec:thaw}): The prior rangefor $\lambda_1 \in [0,\, 2]$. The amplitude $V_1$, related to the present dark energy density is determined by \texttt{CLASS} (shooting) such that the scalar field satisfies the flatness condition. For late time acceleration, the potential must be sufficiently flat. In the case of an exponential potential ($\sim \mathrm{e}^{-\lambda_1 \phi/\Mpl}$), this requirement imposes the condition $\lambda_1 < \sqrt{3}$, allowing the field to thaw slowly and behave as a dark energy component. We therefore adopt the prior range $\lambda_1 \in [0,\, 2]$, which covers this slow roll regime. The limit $\lambda_1 = 0$ corresponds to the $\Lambda$CDM model. Moreover, the thawing dynamics is symmetric under the transformation $\lambda_1 \to -\lambda_1$, so only non negative values need to be sampled. The initial field value at, $\phi_{\rm ini}/\Mpl= 0.2$.

    \item \textbf{Scaling+Thawing} (Subsec.~\ref{sec:sc+thaw}): $\{f_{\rm EDE},\, \log_{10}(1+z_c),\, \lambda_1\} \in [0.001,\, 0.3],\; [3.0,\, 4.3],$ $ [-40,\, 1.5]$.We adopt a large prior of $\lambda_1$ to test this limit. For an exponential potential ($\sim \e^{-\lam_2 \phi/\Mpl}$), the scalar field contribution at this epoch satisfies $3/\lambda_2^2 < \Omega_\phi < 4/\lambda_2^2$. For each point, $\lambda_2$ and $V_2$ are obtained by \texttt{CLASS} (shooting) so that the early field reaches $f_{\rm EDE}$ at $z_c$, and $V_1$ is fixed by the flatness condition,keeping $\phi_{\rm ini}/\Mpl=0.2$ fixed.
    
    \item \textbf{Scaling+$\Lambda$CDM} (Subsec.~\ref{sec:sc+lambda}): $\{f_{\rm EDE},\, \log_{10}(1+z_c)\} \in [0.001,\, 0.3],\; [3.0,\, 4.3]$. As in the Scaling+Thawing case, $\lambda_2$ and $V_2$ are determined numerically by \texttt{CLASS} through a shooting procedure from the sampled values of $f_{\rm EDE}$ and $z_c$, with the initial field value fixed to $\phi_{\rm ini}/\Mpl=1$. The cosmological constant is fixed by the flatness condition.
    
    \item \textbf{Scaling+$w$CDM} (Subsec.~\ref{sec:sc+wcdm}): $\{f_{\rm EDE},\, \log_{10}(1+z_c),\, w_0\} \in [0.001,\, 0.3],\; [3.0,\, 4.3],\; [-3,\, 1]$, with the additional condition $w_0<0$. The early field is treated as in Scaling+$\Lambda$CDM ($\lambda_2$, $V_2$ shot), and the constant-$w$ is fixed by the flatness condition.
    \item \textbf{Scaling+CPL} (Subsec.~\ref{sec:sc+cpl}):
    \begin{align}
        f_{\rm EDE} &\in [0.001,\, 0.3], & \; \log_{10}(1+z_c) &\in [3.0,\, 4.3], \nn \\
        w_0 &\in [-3,\, 1], & \; w_{\rm a} &\in [-3,\, 2], \nn
    \end{align}
    with the additional condition $w_0+w_{\rm a}<0$, which keeps the CPL fluid subdominant at early times. The early field is treated as above ($\lambda_2$, $V_2$ shot).

    \item \textbf{CPL} : $\{w_0,\, w_{\rm a}\} \in [-3,\, 1],\; [-3,\, 2]$, with the additional condition $w_0+w_{\rm a}<0$, which keeps the dark energy subdominant at early times. The cosmological constant is replaced by a fluid with $w(a)=w_0+w_{\rm a}(1-a)$. The priors are identical to those of the CPL fluid in the Scaling+CPL case, so the two models differ only by the early scalar field; $w_0=-1$, $w_{\rm a}=0$ recovers $\Lambda$CDM.
\end{itemize}

\section{Results}
\label{sec:results}
The constraints on the cosmological parameters for the $\Lambda$CDM, Thawing (Sec.~\ref{sec:thaw}), SC$+$TH (Sec.~\ref{sec:sc+thaw}), SC$+\Lambda$ (Sec.~\ref{sec:sc+lambda}), SC$+w$ (Sec.~\ref{sec:sc+wcdm}), SC$+$CPL (Sec.~\ref{sec:sc+cpl}) and CPL models are summarised in Table~\ref{tab1}, and the model comparison statistics in Table~\ref{tab:model_comparison}. For all models the background parameters are tightly constrained and mutually consistent at the $1\sigma$ level, as shown in Fig.~\ref{fig:constraints}. The present matter density lies in the narrow range $\Omega_{\rm m0}\simeq0.302$--$0.310$, with the $\Lambda$CDM value $\Omega_{\rm m0}=0.3023\pm0.0036$ with the models with a CPL component prefer slightly higher values ($\Omega_{\rm m0}=0.3075$ for SC$+$CPL and $0.3102$ for CPL). The Hubble constant is similarly well determined, $H_0\simeq67.7$--$68.3\,{\rm km\,s^{-1}\,Mpc^{-1}}$ in all cases, compared with $H_0=68.26\pm0.28\,{\rm km\,s^{-1}\,Mpc^{-1}}$ in $\Lambda$CDM. None of the models shifts $H_0$ appreciably towards the locally measured value, so none of them alleviates the $H_0$ tension. The physical baryon density is essentially model independent, $\omega_{\rm b}\simeq0.0222$--$0.0224$, and the combination $r_{\rm d}h$, which is directly probed by BAO, is determined at the sub-percent level ($r_{\rm d}h\simeq99.8$--$100.8\,{\rm Mpc}$), again with no significant deviation from $\Lambda$CDM.The angular scale of the sound horizon is fixed by the CMB almost independently of the late time dynamics, $100\,\theta_{\rm s}\simeq1.042$ with an uncertainty of $\sim0.0002$ for all models. The scalar spectral index, $n_{\rm s}\simeq0.964$--$0.970$, the amplitude $\ln(10^{10}A_{\rm s})\simeq3.03$--$3.06$ and the optical depth $\tau_{\rm reio}\simeq0.047$--$0.064$ are also consistent among all models, with the expected positive correlation between $\ln(10^{10}A_{\rm s})$ and $\tau_{\rm reio}$. 
For the scaling models, the data constrain the early dark energy fraction to be very small. We obtain $f_{\rm EDE}\lesssim 0.0075$ and $\log_{10}(1+z_c)$ between $3.31$-$3.65$ for SC$+$TH, SC$+\Lambda$, SC$+w$ while $f_{\rm EDE}\lesssim 0.015$ and $\log_{10}(1+z_c)\approx 3.32$ for SC$+$CPL. Since $\lambda_2$ is determined by $f_{\rm EDE}$, these bounds translate into a steep early potential with the posteriors of $\lambda_2 \gtrsim 30$ for all four models (Table~\ref{tab1} and Fig.~\ref{fig:dark_scf}). This steepness is required for the field to follow a scaling solution, tracking the dominant background component at early times while remaining strongly subdominant \cite{Ramadan:2023ivw}. For an exponential potential $\sim \e^{-\lambda_2\phi/\Mpl}$, the dark energy fraction during the radiation and matter eras is $4/\lambda_2^2$ and $3/\lambda_2^2$, respectively \cite{Copeland:1997et}, so that around matter-radiation equality $3/\lambda_2^2<\Omega_\phi<4/\lambda_2^2$. For $\lambda_2\gtrsim 30$ this gives $\Omega_\phi\lesssim 0.5\%$ at equality. The early dark energy fraction is therefore tightly constrained to be very small for scaling type dynamics, which consequently cannot alleviate the $H_0$ tension. The same argument extends to tracker dynamics, which closely resemble scaling solutions \cite{Steinhardt:1999nw,Zlatev:1998tr,Hossain:2025grx}.

For the late time term, the slope of the single exponential Thawing model is bounded from above, $\lambda_1 < 0.71$, as shown in Fig.~\ref{fig:lambda1_V1} and Fig.~\ref{fig:constraints_lambda}. This is consistent with a nearly flat potential that keeps the equation of state close to $-1$ until recent epochs. In SC$+$TH, negative $\lambda_1$ freezes the field, which then behaves as a cosmological constant, so $\lambda_1$ remains unconstrained. In contrast to the slopes, the late time amplitude of Thawing is well determined, $V_1 = 0.726^{+0.013}_{-0.037}$ in units of $\rho_{\rm c0}$. This is because $V_1$ is fixed by the flatness condition to reproduce the present dark energy density. Since the Thawing field starts from $\phi_{\rm ini}=0.2$ and its prior range on $\lambda_1$ is restricted to small positive values, $V_1$ remains close to $\Omega_{\rm DE0}\simeq0.7$, increasing slightly with $\lambda_1$ (Fig.~\ref{fig:lambda1_V1}). In SC$+$TH, on the other hand, the field is displaced to $\phi_0$ during the early scaling phase, so the flatness condition fixes only the combination $V_1e^{-\lambda_1\phi_0}\simeq\Omega_{\rm DE0}$ and not $\lambda_1$ and $V_1$ individually. Since we have chosen a large negative prior range for $\lambda_1\in[-40,1.5]$, $V_1$ decreases exponentially as $\lambda_1$ becomes more negative and its posterior extends to very low values along this degeneracy as shown in Fig.~\ref{fig:lambda1_V1} and Fig.~\ref{fig:constraints_lambda}. Only for $\lambda_1\to0$ does $V_1$ approach $\simeq0.7$, the value found for Thawing. Therefore, $\lambda_1$ and $V_1$ remain unconstrained in SC$+$TH. The amplitude of the early term $V_2$ is set by $z_c$ through the shooting procedure and therefore inherits the prior dependence of $z_c$.

For the fluid extensions, the SC$+w$ model is fully compatible with a cosmological constant, $w_0 = -1.001 \pm 0.022$. The CPL type models prefer an evolving equation of state with $w_0 = -0.877^{+0.055}_{-0.056}$ and $w_{\rm a} = -0.49 \pm 0.21$ for SC$+$CPL and $w_0 = -0.842^{+0.055}_{-0.050}$ and $w_{\rm a} = -0.63 \pm 0.19$ for CPL. Since $w_0$ and $w_{\rm a}$ are correlated, the significance of this preference is quantified in the two dimensional $(w_0,w_{\rm a})$ space using the likelihood ratio test. Wilks' theorem~\cite{Wilks:1938dza} states that $-\Delta\chi^2_{\rm min}$ follows a $\chi^2$ distribution with $k$ degrees of freedom, where $k$ is the number of additional parameters. For $k=2$ the $p$ value is
\begin{equation}
p=\int_{|\Delta\chi^2_{\rm min}|}^{\infty}\chi^2_{k=2}(x)\,{\rm d}x=e^{-|\Delta\chi^2_{\rm min}|/2},
\label{eq:wilks}
\end{equation}
which corresponds to a Gaussian significance $N\sigma$ through $p={\rm erfc}\left(N/\sqrt{2}\right)$. For CPL, $\Delta\chi^2_{\rm min}=-9.93$ relative to $\Lambda$CDM gives $p=0.0070$, which gives a $2.7\sigma$ preference for evolving dark energy. For SC$+$CPL, the corresponding nested model is SC$+\Lambda$, recovered for $(w_0,w_{\rm a})=(-1,0)$, and $\Delta\chi^2_{\rm min}=-8.79$ gives $p=0.012$ which gives a $2.5\sigma$. This is qualitatively consistent with the preference for evolving dark energy reported by DESI DR2~\cite{DESI:2025zgx,DESI:2025wyn}. The early scaling field leaves these late time constraints essentially unchanged, as the best fit values of $w_0$ and $w_{\rm a}$ in SC$+$CPL and CPL agree within $1\sigma$.

\begin{figure}[t]
\centering
\includegraphics[width=\columnwidth]{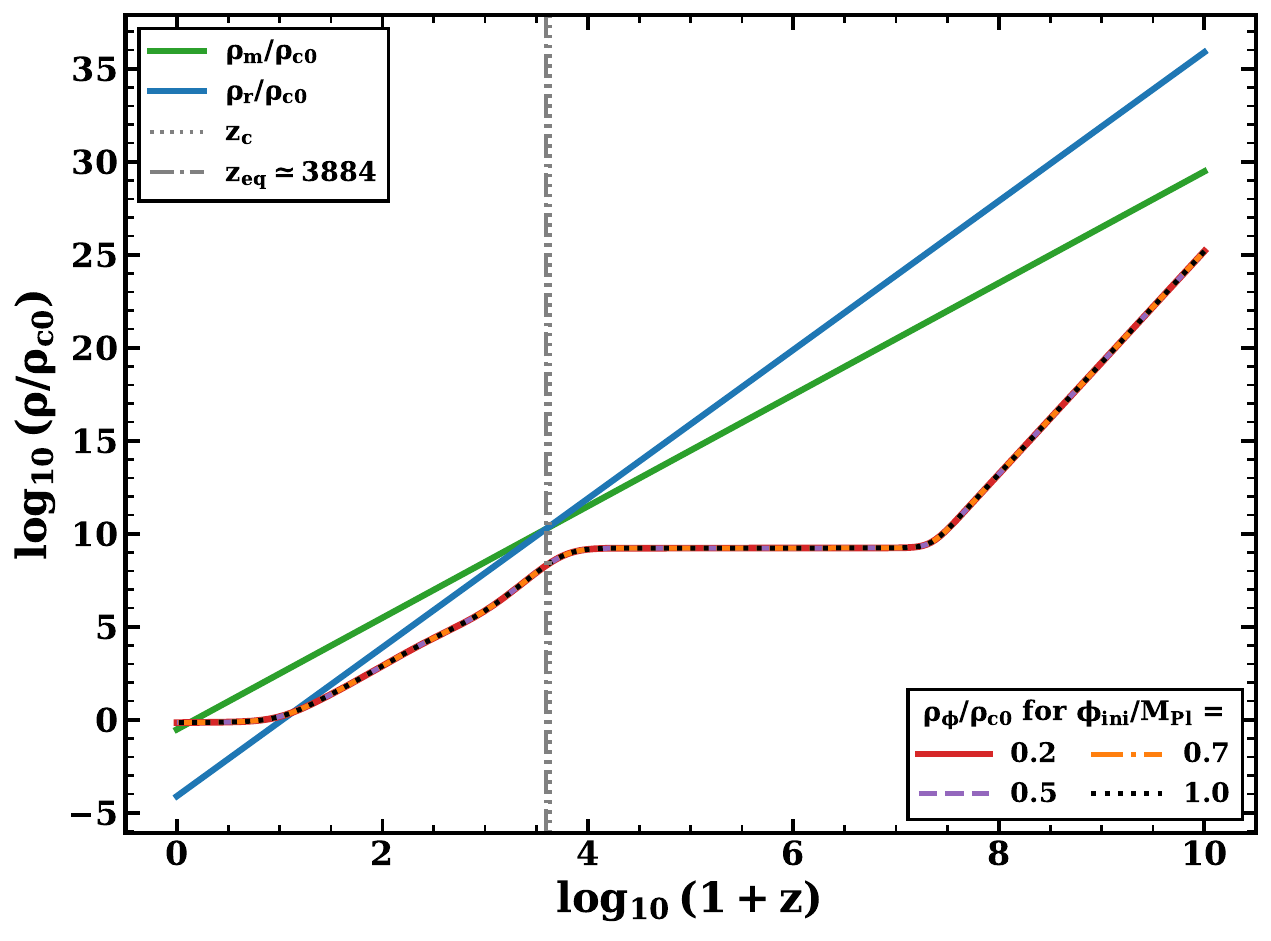}
\caption{Evolution of the matter, radiation and scalar field energy densities in the SC$+$TH model for different initial field values, $\phi_{\rm ini}/M_{\rm Pl}=0.2,\,0.5,\,0.7$ and $1$, with $f_{\rm EDE}=0.005$, $\log_{10}(1+z_c)=3.63$ . The slope $\lambda_2=35.65$ is set by $f_{\rm EDE}$ and is the same for all $\phi_{\rm ini}$, while $V_2$ is obtained by shooting.}
\label{fig:phi_ini_test}
\end{figure}

The redshift $z_c$, at which $f_{\rm EDE}$ reaches its maximum, determines the energy scale at which the frozen field begins to roll. Observational data keeps $z_c$ weakly constrained and prior dominated within our adopted prior range. As shown in Fig.~\ref{fig:phi_ini_test}, different values of $\phi_{\rm ini}$ can produce essentially identical scalar field dynamics, provided that the energy scale associated with the frozen phase, as determined by $z_c$, is kept fixed. In terms of the potential, this energy scale is related to $V_2 e^{-\lambda_2\phi_c/M_{\rm Pl}}$, where $\phi_c \gtrsim \phi_{\rm ini}$ denotes the approximately constant field value during the frozen phase. Thus, fixing $z_c$ effectively fixes the energy scale during this phase. Consequently, for a fixed $\lambda_2$, changing $\phi_{\rm ini}$ modifies the exponential factor and, therefore, requires a corresponding adjustment in $V_2$ to preserve the same energy scale. This explains why similar scalar field dynamics can be obtained for different values of $\phi_{\rm ini}$, as demonstrated in Fig.~\ref{fig:phi_ini_test}. This degeneracy is also reflected in Table~\ref{tab1}, where significantly different values of $\log_{10}V_2$ are obtained for the scaling models. These differences do not correspond to distinct physical behaviours but rather arise from the choice of the initial field value. In other words, for a fixed $z_c$, changing $\phi_{\rm ini}$ effectively rescales $V_2$ while preserving the energy scale associated with the frozen phase.

\begin{figure}[t]
\centering
\includegraphics[width=\columnwidth]{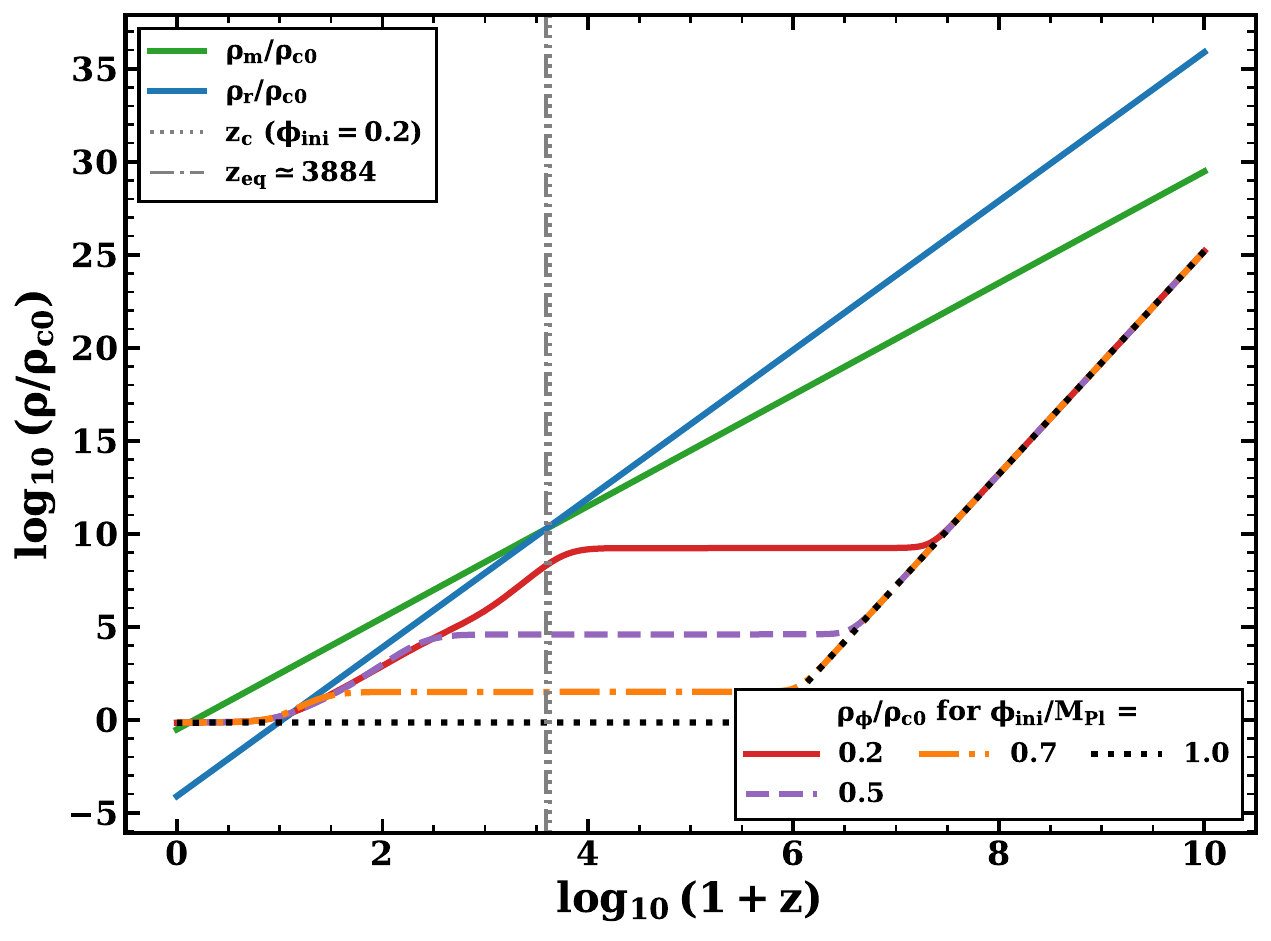}
\caption{Evolution of the matter, radiation and scalar field energy densities in the SC$+$TH model for different initial field values, $\phi_{\rm ini}/M_{\rm Pl}=0.2,\,0.5,\,0.7$ and $1$ without shooting. The slope $\lambda_2=35.65$ and $\log_{10}V_2\simeq14$ are kept fixed . For larger $\phi_{\rm ini}$ the energy scale of the frozen phase decreases, so that $z_c$ and $f_{\rm EDE}$ shift with $\phi_{\rm ini}$.}
\label{fig:phi_ini_fixed}
\end{figure}

On the other hand, if we consider $V_2$ as a model parameter instead of $z_c$, the peak of $f_{\rm EDE}$ is no longer constrained to remain at the same redshift, particularly around matter--radiation equality. This behaviour is illustrated in Fig.~\ref{fig:phi_ini_fixed}, where, for a fixed $V_2$, a larger value of $\phi_{\rm ini}$ lowers the energy scale associated with the frozen phase, causing the field to begin rolling at a later time. Consequently, the onset of the scaling regime is shifted, as $z_c$ now varies with $\phi_{\rm ini}$. Nevertheless, once the field begins to roll, its evolution converges towards the same scaling behaviour, corresponding to the attractor solution.

From the model comparison perspective (Table~\ref{tab:model_comparison}), the CPL models achieve the lowest $\chi^2_{\rm min}$: $\Delta\chi^2_{\rm min} = -9.93$ for CPL and $-7.62$ for SC$+$CPL. For CPL, this improvement survives the penalty for its two additional parameters under AIC and DIC ($\Delta{\rm AIC}=-5.93$, $\Delta{\rm DIC}=-5.98$), indicating a preference for CPL over the reference model. However, the stronger complexity penalty imposed by BIC disfavors CPL ($\Delta{\rm BIC}=8.82$). A direct comparison of SC$+$CPL with CPL for the same datasets, obtained from the values in Table~\ref{tab:model_comparison}, shows that adding the early scaling field does not improve the fit, with $\Delta\chi^2_{\rm min}=+2.31$, $\Delta{\rm AIC}=+6.31$ and $\Delta{\rm DIC}=+2.54$ relative to CPL. Relative to $\Lambda$CDM, its two additional scaling parameters raise the AIC to the $\Lambda$CDM level ($\Delta{\rm AIC} = 0.38$), although the DIC, which accounts only for the parameters actually constrained by the data, still mildly favours it ($\Delta{\rm DIC} = -3.43$).The Thawing model gives a small improvement in fit ($\Delta\chi^2_{\rm min} = -1.64$) but no preference once its extra parameter is penalised ($\Delta{\rm AIC} = 0.36$, $\Delta{\rm DIC} = -0.29$). The scaling models with a cosmological constant, a constant $w$ or a thawing late term fit the data as well as $\Lambda$CDM, with $\Delta\chi^2_{\rm min}$ between $0.1$ and $1.4$. Since these models contain $\Lambda$CDM as a limit, these small positive values reflect the finite precision of the best fit estimated from the chains rather than a worse fit. They are disfavoured by all criteria ($\Delta{\rm AIC} \simeq 5$--$7$, $\Delta{\rm BIC} \simeq 20$--$30$).
Overall, the data show a mild preference for late time evolving dark energy, the requirement of an early time scaling regime suppresses this preference.

\section{Conclusions}
\label{sec:conc}

In this work we have investigated a broad class of dynamical dark energy scenarios, including thawing, scaling+thawing and effective fluid extensions (SubSecs.~\ref{sec:thaw}--\ref{sec:sc+cpl}) with Planck PR4 CMB temperature, polarisation and lensing data, DESI DR2 BAO and Pantheon+ supernovae, using a full Boltzmann treatment with \texttt{CLASS} and \texttt{Cobaya}. We find that the standard background parameters are tightly constrained and remarkably stable across all models. In particular, $\Omega_{\rm m0}$, $H_0$, $\omega_{\rm b}$ and $r_{\rm d}h$ are mutually consistent at the $1\sigma$ level, with $H_0 \simeq 67.7$--$68.3\,{\rm km\,s^{-1}\,Mpc^{-1}}$ in all cases, as summarised in Table~\ref{tab1} and illustrated in Fig.~\ref{fig:constraints}. Neither thawing nor scaling dynamics therefore induces a statistically significant modification of the background expansion, and none of the models alleviates the $H_0$ tension.

At late times, the models with a dynamical equation of state described by the CPL parametrization prefer a departure from a pure cosmological constant. For CPL we find $w_0 = -0.842^{+0.055}_{-0.050}$ and $w_{\rm a} = -0.63 \pm 0.19$, corresponding to a $\sim 2.7\sigma$ preference over $\Lambda$CDM ($\Delta\chi^2_{\rm min} = -9.93$). A key outcome of our analysis concerns the interplay between late time dark energy dynamics and early time behaviour. At late times, models allowing for a dynamical EoS, most notably the CPL parametrization discussed in Sec.~\ref{sec:sc+cpl}, exhibit a ($\sim2\sigma$) preference for evolution away from a pure cosmological constant, driven primarily by phantom like behaviour at low redshift. This is the behaviour reported by DESI DR2: $w_0 > -1$ today with $w_{\rm a} < 0$, so that the equation of state was phantom like in the past and crosses $w = -1$ at low redshift. In contrast, the constant-$w$ and thawing models remain fully consistent with a cosmological constant ($w_0 = -1.001 \pm 0.022$ for SC$+w$, and $\lambda_1 < 0.71$ for Thaw).

When an early scaling phase is included (SubSecs.~\ref{sec:sc+thaw}, \ref{sec:sc+lambda}, \ref{sec:sc+wcdm} and \ref{sec:sc+cpl}), the data constrain the early dark energy fraction to be very small, $f_{\rm EDE} < 0.0075$--$0.015$. This requires a steep early potential, $\lambda_2 \gtrsim 30$, so that the scalar field energy density remains strongly suppressed during radiation and matter domination, with $\Omega_\phi \lesssim 0.5\%$ around matter-radiation equality. Scaling type dynamics being tightly constrained, are unable to support a significant early dark energy component and leave the sound horizon and the inferred $H_0$ essentially unchanged.

From a model comparison perspective (Table~\ref{tab:model_comparison}), CPL is favoured by the AIC and the DIC ($\Delta{\rm AIC} = -5.93$, $\Delta{\rm DIC} = -5.98$), indicating positive support for late-time evolving dark energy, while the BIC, with its stronger penalty on additional parameters, disfavours it ($\Delta{\rm BIC} = 8.82$). Adding an early scaling field never improves the fit. The scaling models with a cosmological constant, a constant $w$ or a thawing late term fit the data as well as $\Lambda$CDM but are disfavoured by all criteria once their additional parameters are penalised ($\Delta{\rm AIC} \simeq 5$--$7$, $\Delta{\rm BIC} \simeq 20$--$30$). For SC$+$CPL, the improvement in fit comes entirely from its CPL component, and the scaling parameters only increase the AIC to the $\Lambda$CDM level ($\Delta{\rm AIC} = 0.38$). The Thawing model gives a small improvement in fit but no preference once its extra parameter is taken into account.

Overall, our results indicate that while late time phantom or evolving dark energy may be favoured by current observations, scenarios in which this behaviour is embedded within a framework featuring simple early time scaling are disfavoured. Viable extensions of $\Lambda$CDM that aim to accommodate both early and late time dark energy dynamics therefore require departures from straightforward scaling behaviour or the introduction of additional physical mechanisms beyond those explored in this work.

\section*{Acknowledgments}

The authors acknowledge the High Performance Computing facility Pegasus at IUCAA, Pune, India, for providing computing resources. SA acknowledges the financial support received from the Council of Scientific and Industrial Research (CSIR), Government of India, through the CSIR NET-SRF fellowship (File No. 09/0466(12904)/2021).

\bibliographystyle{apsrev4-2}
\bibliography{references}
\end{document}